\documentclass[oneside,a4paper]{article}

\usepackage{hyperref}
\usepackage{authblk}
\providecommand{\keywords}[1]{\textbf{\textit{Keywords:}} #1}

\title{Giving the Right Answer: a Brief Overview on How to Extend Ranking and Skyline Queries}
\author{Sergio, SC, Cuzzucoli}
\affil{Politecnico di Milano\\
Milan, Italy\\
\href{mailto:sergio.cuzzucoli@mail.polimi.it}{sergio.cuzzucoli@mail.polimi.it} }
\date{}

\begin{document}
\maketitle
\begin{abstract}

To retrieve the best results in a database we use Top-K queries and Skyline queries but some problems arise. The formers rely too much on user preferences, which are difficult 
to quantify and may skew the fetching of the data, while the latters tend to output too much data. In this paper, we explore three different branches of research that seek to 
overcome such limitations:
Flexible/Restricted Skylines, Skyline Ordering/Ranking, and Regret Minimization. We analyze how they work and we make comparisons among them to guide the reader to choose the
approach that best fits their use cases.

\end{abstract}

\keywords{top-k, Skyline, query, flexible, ordering, ranking, regret, minimization}

\section{Introduction}

In the last years databases have dramatically increased in size. Since users want to find useful information in them, this has led to the problem of understanding how
to actually output the most interesting records when queried.
Think, for example, of a database $r$, populated by $N$ entries ($e_i$ with $i \epsilon [1,N]$) described by M attributes each ($m_i$ with $i \epsilon [1,M]$).
\newline
\newline
One of the simplest approaches to return the most interesting tuples is by using \textbf{top-k queries} (also called \textbf{ranking queries}).
In this case a m-objective problem is simplified into a single objective one: the user provides a list of weight vector $w=< w_1, w_2 ...., w_m>$,  
a score is then computed via the use a scoring function \textit{f} ( one of the simplest ones is the inner product $f(e_i)=\sum_{i=1}^{M}{w_i * k_i}$ ) for every entry. 
Out of all the entries, the first $k$ ones with the highest scores are retrieved.
Another alternative are \textbf{Skyline queries}. They revolve around the idea of \textit{dominance among points}:
given two entries (here called points) $p_1$ and $p_2$, $p_1$ is said to dominate $p_2$ ($p_1 \prec p_2$) iff $p_1$ is not worse than
$p_2$ in any attribute and better than $p_2$ in at least one of them. 
In this case, the desired results are contained in the \textit{Skyline} (SKY(R)), that is the set of points not dominated by any other point.
\newline
\newline
The idea behind top-k queries is that the weight offered by the user are personal, thus the result list is expected to contain the points which are of actual interest to them.
The size of the result list can be easily controlled, making it manageable for the user to check and select its preferred entry.
Unfortunately, these queries are very dependent on the values of user preferences (a slight change in the weights leads to totally different result) and these are generally very 
difficult to quantify. Since they do not offer a global view of the database, they may exclude feasible points because the user has input the "wrong" values. 
The user could expand the pool of tuples simply trying different combinations of weights and hoping to find the best result eventually but this is not only expensive in terms of 
time, it does not guarantee success.
On the other hand, Skyline queries do not require a weight list nor a scoring function, leaving user preferences out of the equation. They assure that there is no point not in 
the Skyline preferrable to the output since every point not in the Skyline is outperformed by at least one point in it. 
Their main drawback is that they are unable to control the size of the result list which grows especially in case of high dimensionality or when data is anticorrelated.
Returning a considerable chunk of the database and expecting the user to easily pick up their preferred point is not ideal. 
\newline
\newline
Due to these limitations numerous new studies have been conducted and they have led to the creation of new procedures and algorithms which seek to overcome them.
In this survey we will focus on those techniques which fall under three different categories, namely those based on \textbf{Flexible and/or Restricted Skylines}, 
\textbf{Skyline Ordering and/or Ranking}, and \textbf{Regret Minimization}.
The rest of the paper is organized as follows: Section 2 is dedicated to explaining flexible and/or restricted Skylines and their underlying concepts, Section 3
focuses on Skyline ordering and/or ranking, Section 4 revolves around the regret minimization approach and its expansions, in Section 5 a comparison is made
among all these approaches and their pros and cons are evalutated, while Section 6 contains all the final considerations of the paper.


\section{Flexible and/or Restricted Skylines}

All the techniques contained in this section try to suppress or mitigate the shortcomings of the methods we have already mentioned, simply generalizing them or 
improving their flexibility via the introduction of new operators, which may also combine properties from top-k and Skyline queries. 

\subsection{Flexible Skylines}
The \textbf{R-Skyline}, first introduced in \cite{DBLP:journals/pvldb/CiacciaM17}, stands on the idea of \textsc{F-dominance}:
given two points \textit{$p_1$} and \textit{$p_2$} and a family of scoring functions \textbf{$F \subset M$}, where $M$ is the family of Monotonone Functions (functions which are always either
increasing or decreasing), $p_1$ F-dominates $p_2$ ($p_1 \prec_F p_2$) when the score of $p_1$ is equal to or better than that $p_2$ using every function contained in F. 
R-Skylines output two non-dominated restricted Skyline of r with respect to F ($ND(r,F)$), which contains all the tuples non-F-dominated in the database r, and the potentially
optimal restricted Skyline of r with respect to F, $PO(r,F)$, populated by the Potentially Optimal Tuples, namely those tuples which are part of the Skyline of at least one of the 
functions of the $F$ family.
Since $F$ can include not only finite but also infinite set of scoring function, as highlighted in \cite{DBLP:journals/tods/CiacciaM20}, it allows to express more general constraints
w.r.t top-k queries but which are indeed powerful in filtering unwanted tuples. Think of someone who wants to buy a new house in Milan, choosing among the values of a database which lists
also the distance between the house and the Linate Airport and the distance to the nearest subway station, both expressed in metres.
With the flexible Skylines just presented, they would be able to express a constraint $F=\{w_S*Subway + w_L*Linate| w_M \ge w_C\}$ which states that they 
prefer to have a station nearby than to live near the airport.
This is one example of a very simple constraint, but the framework allows even more precise ones such as "attribute $a_1$ is more important that $a_2$ but not three times as important",
so the user in question could be able to tweak their choices in order to better balance these two values.
Anyway, the results retrieved using these techniques are obviously of lower cardinality than those of the Skyline as they filter some uninteresting points, in this case homes that could
be greater in size but that are not well served by underground transportation. 
The problems which are tied to the insertion of personal preferences are still included but at least they are mitigated as the algorithms allows to implement general more general constraints
than those used by Top-K queries.
This paper does not intend to examine in depth the implementation of these methods, so the reader who wants to inform on the algorithms used to implement R-Skyline is advised to check
\cite{DBLP:journals/pvldb/CiacciaM17}\cite{DBLP:conf/sebd/CiacciaM18}\cite{DBLP:journals/tods/CiacciaM20}
The same authors have also offered a way to rank object in case of multiple sources and partially specified weights, described by the use of 
FSA in \cite{DBLP:conf/cikm/CiacciaM18}. FSA is based on Fagin's Algorithm (FA) and the Treshold Algorithm (TA) but, differently from them, it requires that weights not be explicitly
specified but rather expressed via a set of constraints, using the concept of F-dominance already seen and proving its power in more than a single set of problems.

\subsection{Uncertain Top-K}
Another way to tackle the problem is given by the Uncertain Top-K Query (UTK), described in \cite{DBLP:journals/pvldb/MouratidisT18}, which deals with the uncertainty in the definition 
of a Top-K Query using two different version: $UTK_1$ and $UTK_2$. $UTK_1$ reports all the tuples that rank among the top-k in a certain region R delimited by an approximate description
of the weight values. In this case the result is minimal, meaning that the user is returned a set which contains points that belong to the top-k set for at least one weight vector in R.
$UTK_2$ expands these results by reporting the top-k for every possible weight vector that lies inside the region, practically returning a partition of R which specifies the Top-K Query for 
every sub-region. These results allow the user to inspect the top-k results for similar preferences.
The novelty of this approach is that it does not deal with uncertain data but with uncertain queries and there are no probabilities affecting its output.
Using this method the user gets a broader view of the database w.r.t that offered by simple Top-K Query, thus they suffer less backlash in the choice of the weights.
In addition, $UTK_1$ inherits from ranking queries the convenience of knowing the cardinality of the results aforehand.

\subsection{Trade-offs in Skyline}
One of the reasons one prefers Skyline queries to Top-K ones is that they do not need user preferences, which are unfeasible to quantify precisely. In \cite{DBLP:conf/edbt/LofiGB10}, the authors discuss the
introduction of trade-off (and user preferences conversely) to enhanche Skyline queries. The reasoning behind these choice is that users tend to compromise on some attributes of a given object 
in order to obtain some other values, a nuance which is not considered when Skylines are computed. In fairness, these are personal choice and, in the examples of the paper, impossible to estimate
without external inputs. This is why \cite{DBLP:conf/edbt/LofiGB10} requests that users indicate their Trade-off are inserted in the equation. For example, let us suppose we have the database
of an e-commerce site which sells only smartphones, of which we will be discussing only the attributes \textit{cost} and \textit{OS} for the sake of simplicity. The user who inputs the tradeoff 
$t_1=((1200, AppearOS) \triangleright (1000, RobotaOS))$ is saying that they would rather prefer Phone A, equipped with AppearOS, even tough it costs 1200\$ , to Phone B that is 1000\$ but that
does not have RobotaOS, as a result (Happear8, 1200, AppearOS) now dominates (SunPhone, 1000, RobotaOS) even tough the opposite would be true using Skylines.
This is only a simple example and in the paper the authors explain how to chain more examples to form more complicated ones.
It is trivial to say that these choice is not dictated by some hard mathematical law, the reason someone preferes an OS or a feature over other qualities are not written in stone. 
That is where the strength of this paradigm lies: in its ability to capture nuances that would otherwise be lost and without which the results would be different.
Obviously this method retains some problems, not only there is no explicit limitation on the size of the output, but there is still the problem of correctly assessing user preferences.


\section{Skyline Ordering/Ranking}

Ranking Skyline points in order to allow the user to retrieve those that they would feel most confident about is another branch of research that has interested scholars who want
to overcome the limitations of Skylines. As the name suggest, the set of techniques here presented combine concepts of Skyline and Top-K Query to offer a better result to the users.

\subsection{Top-K Dominating}
One of the simplest examples of this is Top-K Dominating Queries \cite{DBLP:conf/vldb/YiuM07}, that return an output made of K elements which rank the best according to a scoring function, 
just like any Top-K Query would do. The elements in question are Skyline points ranked using the function ($\mu(p)$)  which computes the score for every point based on the number of points 
dominated by it.

\subsection{Skyrank}
A similar concept is exploited in \cite{DBLP:journals/dke/VlachouV10}, ranking different points using the idea of subspace domination. Subspaces are those spaces
generated by selecting only a part of the dimensions of the database, so it is entirely possible for point $p_1$ to dominate point $p_2$ only on one subspace out of all the combinations. 
SKYRANK is a link-based algorithm which uses the subspace-dominance metric to assign a weight to every point in the Skyline. 
What this does is it creates a graph which shows the (subspace-)dominance, giving the heaviest weight to those point which dominate the most points in different subspaces,
favoring them over \textit{extreme points}, AKA the Skyline points of the 1-dimentional subspaces, which would normally appear using Skylines.
The strength of this approach is that it allows the user to have a balanced view of the database, pruning those points that outperform others only on a single dimension,
it returns intersting point user and it reduces the number of tuples returned to them, while simultaneously not requiring a preference function.
In addition, the authors point out that the algorithm can be extended to accomodate Top-K queries.

\subsection{Size Constrained Skyline}

When we talk about Skyline computation, the first problem we think of is the excessive number of tuples it returns,the authors of \cite{DBLP:journals/tkde/LuJZ11} argue that the
inability to control the size of the result causes problems even in the case where the size of the result list is too low.
For example, if someone wants to buy five used vans for a small shipping company with a low budget, they will not be happy if the e-commerce they use returns only a Skyline of three vans.
The algorithm we are about to discuss solves this problem merging the \textit{pointwise approach}, which is giving a score to the points and sorting them, and the \textit{set-wide maximization approach}, 
which considers a group of points, instead of single ones, and maximizes a target value.
In this case all the data is partitioned into $S = \{ S_1,S_2, ... , S_n \} $ partitions, where $S_1$ is the Skyline of the database, every point of $S_{i+1}$ is dominated by at least one point
in $S_{i}$ (the viceversa does not hold), and the union of all the $S_i$ return the entire database.
Put it simply, the way this is formed is by computing the Skyline of the database and removing the points contained in the Skyline, iteratively, until we are left with a layered structure
that contains all the points of the database.
When a user requests the best $k$ tuples of the database, the system proceeds to output them according to the scores computed using the scoring function of choice ( e.g the number
of dominated points or subspace Skyline frequency). If $k > |S_1|$, the method outputs all the points from $S_1$ then goes to the next layer, $S_2$, and so on, until $k$ elements are returned.

\subsection{Skyline Ordering Using Partially Ordered Domains}

So far we assumed that, when ordering Skylines, we were able to establish an order for every attribute we were discussing (e.g lower the price, increase the horse power). 
In reality is not possible to outline an univocal order for every kind of attribute present on the database, simply because many times there is not a value which is
objectively better than another one, such as in the case of \textit{nominal attributes}. An Appear phone is not preferrable to a RobotaOS phone per se but a user could reason this way.
This is often not discussed: user preferences can be used to sort the points inside a Skyline, asking the user if they prefer a precise value
over another allows us to further filter the results to give back to them, considerably reducing the size of the output. This is the intuition behind \cite{DBLP:journals/tkde/WongPFW09}. 
In this paper, the Skyline Query is contained in the Ordered Skyline Tree, or better in its compressed form which is the CST or Compressed Ordered Skyline Tree. 
What this structure does is it splits the Skyline into the \textit{global Skyline point set} ($G$) and \textit{the order-sensitive Skyline point set} ($D'$), basically precomputing how to return
different results to different users with different tastes or interests.
The former set contains all the point which are part of the Skyline in any case, the latter is susceptible to user choices. 
In the work examined, such choices are expressed as partial orderings, for example the user who prefers an Appear phone over a RobotaOS one one will impose $AppearOS \ge RobotaOS$ and they will get an output
which contains Appear phone, filtering the RobotaOS ones they are not interested in.
The peculiarity of this approach is that it exploits an often overlooked set of attributes and it mitigates the problems strictly tied to Skyline queries.
The word "mitigate" here is not used casually, unfortunately the resulting tuples can still be high in number as there is no hard constraints on their cardinality.
Suppose that Appear is considered a luxury brand and, even tough it has a considerable share of the marketplace, its phones still are far below the 50\% of the pool of the
smartphones in circulation, the other ones mounting RobotaOS; if a user imposes the constraint "RobotaOS over AppearOS", they will still be inundated with results if no other constraints are
imposed. This approach improves with more constraints.


\section{Regret Minimization}

The regret minimization query is an approach first introduced in \cite{DBLP:journals/pvldb/NanongkaiSLLX10}. Differently from what we have seen so far, its objective is not to maximize the
satisfaction of the user but rather to minimize their dissatisfaction. Similarly to Ranking Queries, it returns a fixed amount of tuples and, as Skyline Queries do, it does not require
to input a precise scoring function. Via the use of the \textit{k-regret operator}, the k-regrets query returns a representative set of the database, composed of k elements, which aims to
minimize the value of the maximum regret ratio of the user, where the regret ratio is the percentage of dissatisfaction they would have choosing one point belonging to the result instead 
of the best single point present in the database. This hypothesis is relaxed in \cite{DBLP:journals/pvldb/ChesterTVW14} where the regret ratio is computed using the result of a Top-K Query
run on the database as a comparison, instead of the best single point.
In any case the k-regret operator is \textit{scale invariant}, thus the scale of the dimensions of the database does not change the solution
(the result does not change if an attribute is expressed in meters or in miles), and it is \textit{stable}, which means that the addition of \textit{junk points}, points which are not of
interest to the user, has no effect on the final results. The stability of the operator could be a much appreciated property, for example in the case of the database of an e-commerce
open to every seller, as it avoids that someone games the algorithm used by the system by placing a great deal of knock-off objects hoping to influence the results returned to the user. 
Actually, the results are influenced by the choice of the \textit{utility function} (a synonym for the scoring funcion), as it mimics the behavior of the happiness of the user.
Since there is no way to know the which utility function to use beforehand, many people default to using linear utility ones, see for example \cite{DBLP:journals/pvldb/NanongkaiSLLX10}, 
\cite{DBLP:conf/icde/ZeighamiW19}, 
\cite{DBLP:conf/sigmod/NanongkaiLSM12}. This choice is dictated by the fact that not only this is one of the simplest class of functions, but it models efficiently the behavior of the user
in most of the cases (the same intuition behind Top-K Queries). However, the question on whether linear ones are the best among the utility functions is not trivial.
Falkner, Brackenbury and Lall \cite{DBLP:journals/pvldb/FaulknerBL15} capture a nuance which can't be modeled by linear functions, which is \textit{diminishing marginal returns}, or the
idea that the user loses interest in the increase of the value of an attribute as it grows. For example, suppose that you want to buy a new camera and that you can choose
between Camera A with 12 Megapixels and Camera B with B with 22 Megapixels, if you value Megapixels, you will surely prefer B over A by a large margin. Suppose now that you have yet another
choice which is Camera C with 25 Megapixels: probably you will prefer C over B but less strongly than how you prefer B over A, as the last choice is between two cameras that already have many
Megapixels. That is an example of reasoning behind the proposal for the use of concave, convex or Constant Elasticity of Substitution (CES) functions, further detailed in 
\cite{DBLP:journals/tods/QiZSY18}, that allow to better model these types of phenomena.
Another important aspect of the regret minimization problem revolves on what metric to minimize. As already stated, \cite{DBLP:journals/pvldb/NanongkaiSLLX10} proposes to reduce the maximum
regret ratio as much as possible, but that is not the only option. The authors in \cite{DBLP:conf/icde/ZeighamiW19} argue that this choice produces a worst-case guarantee on the
regret ratio of the user, but it is not effective in keeping low the regret ratio for the average user, saying that it should be the priority, thus they provide a framework where the problem is 
formulated using the average regret ratio as a metric. Still, these are both particular cases, so in \cite{DBLP:journals/pvldb/ShetiyaAAD19} a unified algorithm is produced which, by the
use of "oracles", decides which of the two metrics to use for the problem at hand.
Finally, it is possible to relax also the assumption that to minimize the regret ratio of the user no interaction is needed. In \cite{DBLP:conf/sigmod/NanongkaiLSM12}, the regret minimization
algorithm is enhanched with \textit{interaction} with the user, obtained by asking the user questions before the computation of the results. This questions are asked in rounds and they consist
in requiring the user to pick a tuple between a set. The tuples can also not even be in the database, as they are used just to understand how the users weighs the different dimensions
of the element, but they should be at least representative of the actual points, least there is the risk of disappointment in the result. The user can decide to stop whenever he/she wants
and, when it happens, be prompted with the results. The authors claim that only a small set of questions is necessary to improve the results so even tough the user has to put some effort in the 
research, at least the task is not cumbersome.


\section{Comparison}

\begin{table}
    \caption{A brief recap of the properties of the techniques}
    \centering
    \begin{tabular}{| l | c | c | c |}
      \hline
      Property           & Flexible/Restricted Skylines & Skyline Ordering & Regret Minimization \\
      \hline
      User Interaction   & Yes                          & Possible         & Possible \\
      \hline
      Output Cardinality & Reduced/Fixed                & Reduced/Fixed    & Fixed \\
      \hline
      Scale Invariance   &                              & Yes              & Yes\\
      \hline
      Stability          & Yes                          & Yes              & Yes \\
      \hline
    \end{tabular}
    \label{tbl:recap}
  \end{table}

We have different approaches to improve the basic queries used to satisfy the user, but since each one of them is different from the others and it carries its own 
benefits and drawbacks, we are now going to draw a comparison to better understand when to use them.
As pointed in \ref{tbl:recap}, generally flexible and/or restricted Skyline are best exploited when the user is able to personalize their research. This kind of queries 
allows to correlate different attributes either by imposing constraints in order to 
to use R-Skylines, or to define some weights to compute Uncertain Top-K Query, without having to worry to much about having to tinker their values to get the best results, or to impose 
some trade-offs
to shape a better Skyline. In the first two cases it is also possible to accurately limit the cardinality of the results thus we are sure not to overwhelm the user with excessive
information.
In case we want to ease (or outright eliminate) user interaction while still limiting the number of tuples returned from the system, one of the best choices is surely using Skyline Ordering.
In the first scenario, we can still have a fixed amount of elements output, if we rely on Top-K Dominating Queries, Skyrank or Size Constrained Skyline Queries, without even asking the user
for interaction.
On the other hand, if we accept to not be able to predict exactly how many elements the user will have to inspect in order to pick its favorite, we can exploit nominal values while keeping
the global view of the Skyline, at the cost of some simple constraints to be imposed by the user, a much needed feature when we have a database filled with information that can't be ordered
otherwise.
Using Regret Minimization techniques we have to withstand some uncertainty in the results but we are able to better predict how much the user will be interested in a set of fixed values
while not having to even define a precise scoring function in the process. Furthermore, if we want to reduce said uncertainty, we can just enable some basic user interaction to refine the 
results.


\section{Conclusions}

In this paper we have review how to go beyond the use of Top-K and Skyline queries in order to avoid the complications and drawbacks they carry. 
We have done so by splitting our analysis into three different families of approaches and outling their pros and cons, finally we have tried to outline also some scenarios in which to 
use one approach or the other. We came to the conclusion that it is best to use Flexible and/or Restricted Skylines when there is no problem in asking the user their scoring function.
If the user has partial knowledge of their preferences and they are willing to pick a point among those we give them, ordering the Skyline is acceptable. 
In the case where the user does not know how to impose an order, but they can possibly accept some minor interaction and some imprecision in the result, regret minimization is possible.

\bibliographystyle{plain}
\bibliography{refs.bib}
\end{document}